\newcommand{\micron}{$\mu$m}
\newcommand{\lya}{Ly$\alpha$}
\newcommand{\halpha}{H$\alpha$}

\newcommand{\lha}{$L_{\mathrm{H}\alpha}$}
\newcommand{\lfha}{$LF($H$\alpha)$}
\newcommand{\ewha}{$EW_{\mathrm{H}\alpha}$}
\newcommand{\sfrd}{$\dot{\rho}_{\star}$}
\newcommand{\lumd}{$\rho_L$}
\newcommand{\ks}{${K_\mathrm{s}}$}

\newcommand{\ebv}{$E_{B-V}$}
\newcommand{\av}{$A_V$}

\newcommand{\lstar}{$L_\star$}
\newcommand{\phistar}{$\phi_\star$}

\newcommand{\lumcgs}{erg~s$^{-1}$}
\newcommand{\permpc}{Mpc$^{-3}$}
\newcommand{\sfrvol}{$M_{\sun}$~yr$^{-1}$~Mpc$^{-3}$}

\documentclass[letter,traditabstract]{aa} 
\usepackage[]{natbib}
\usepackage{graphicx}
\usepackage{txfonts}

\begin{document}
\title{The H-alpha luminosity function at redshift 2.2 
\thanks{Based on observations made with ESO Telescopes at the 
Paranal Observatory under programme ID 081.A-0932}}
\subtitle{A new determination using VLT/HAWK-I}

\author{Matthew Hayes
\inst{1}
\and
Daniel Schaerer
\inst{1,2}
\and
G\"oran \"Ostlin
\inst{3}
}

\institute{Geneva Observatory, University of Geneva, 51 chemin des Maillettes, 
1290 Sauverny, Switzerland\\
\email{matthew.hayes@unige.ch} 
\and
Laboratoire d'Astrophysique de Toulouse-Tarbes, Universit\'e de Toulouse, CNRS, 
14 Avenue E. Belin, 31400 Toulouse, France 
\and 
Oskar Klein Center for Cosmoparticle physics, Department of Astronomy, 
Stockholm University, 10691 Stockholm, Sweden
}

\date{Received September 01, 2009; accepted December 07, 2009}

\abstract{ We aim to place new, strengthened constraints on the luminosity 
function ($LF$) of H-alpha (\halpha) emitting galaxies at redshift 
$z\approx 2.2$, and to further constrain the instantaneous star-formation rate 
density of the universe (\sfrd).
We have used the new \emph{HAWK-I} instrument at \emph{ESO-VLT} to obtain
extremely deep narrow-band (line; \emph{NB2090}) and broad-band (continuum; \ks) 
imaging observations. 
The target field is in the GOODS-South, providing us with a rich 
multi-wavelength auxiliary data set, which we utilise for redshift confirmation
and to estimate dust content. 
We use this new data to measure the faint-end slope ($\alpha$) of \lfha\ with 
unprecedented precision.
The data are well fit by a Schechter function and also a single power-law, 
yielding $\alpha = (-1.72\pm 0.20)$ and $(-1.77\pm 0.21)$, respectively. 
Thus we are able to confirm the steepening of $\alpha$ from low- to high-$z$ 
predicted by a number of authors and observed at other wavelengths.
We combine our $LF$ data-points with those from a much shallower but wider 
survey at 
$z\sim 2.2$ (Geach et al. 2008), constructing a $LF$ spanning a factor 
of 50 in luminosity.
Re-fitting the Schechter parameters, we obtain 
$\log L_\star    = (43.07\pm  0.22)$~\lumcgs;
$\log \phi_\star = (-3.45\pm  0.52)$~\permpc;
$\alpha          = (-1.60\pm  0.15)$. 
We integrate over \lfha\ and apply a correction for dust attenuation
to determine the instantaneous cosmic star-formation 
rate density at $z\sim 2$ without assuming $\alpha$ or extrapolating it from 
lower-$z$.
Our measurement of \sfrd\ is $(0.215\pm 0.090)$~\sfrvol, integrated over a range of 
$37\le \log (L_{\mathrm{H}\alpha} /\mathrm{erg}~\mathrm{s}^{-1})\le 47$.  }

\keywords{Galaxies: evolution -- Galaxies: high-redshift  -- Galaxies: starburst }
\maketitle

\section{Introduction}

Since the evolution in the rate of cosmic star-formation was first plotted 
\citep{Lilly1996,Madau1996}, 
the literature has been awash with studies adding further points to the diagram.
To estimate this quantity, one typically needs to find a number of galaxies by 
a given method and  estimate their density in both 
space and luminosity (i.e. the luminosity function, $LF$).
Integration of $LF \cdot L \cdot dL$ thus provides the volume-averaged 
luminosity density (\lumd) and, if $L$ is a suitably calibrated measure of 
star-formation rate (SFR), \lumd\ converts directly into
the volumetric rate of star-formation (\sfrd).

A favourite tracer among low-$z$ observers for estimating star-formation 
is the \halpha\ emission line due to its simple physics, 
high intrinsic brightness, and convenient rest-wavelength at 6563\AA.
At $z \gtrsim 2.5$, however, the \halpha\ line shifts out of the $K-$band,
making it highly inefficient for galaxy evolution studies. Fortunately
at this $z$, selection by either the \lya\ line or continuum
($BM/BX$, $BzK$) criteria becomes possible, but unfortunately a different 
population of galaxies may be recovered, introducing biases in selected 
star-formation rate/history, dust content, etc.
More specifically with regard to SFR, continuum luminosities come to equilibrium
(and are therefore calibrated) over very different time-scales to nebular lines:
$\gtrsim 100$ and $\sim 10$~Myr respectively, implying that lines are more 
sensitive to the instantaneous SFR
\citep[e.g.][]{Kennicutt1998}. 
\lya\ enables surveys to go much further in redshift and has 
identical production physics, but is a resonance line and undergoes a complex 
radiation transport, which renders it an unreliable tracer of intrinsic properties. 
From a purely physical perspective, \halpha\ is a far preferable tool.

Luminosity functions are typically parameterised and compared using the Schechter 
function\footnote{$\Phi(L)\cdot \mathrm{d} L = \phi_\star \cdot (L/L_\star)^{\alpha} \cdot 
	\exp(L/L_\star)  \cdot \mathrm{d}L/L_\star$}
in which the luminosity distribution below (above) the 
characteristic luminosity, \lstar, is dominated by a power-law
(exponential). 
When fitting, the three Schechter parameters are degenerate,
and strong constraints are only obtained by sampling above and 
significantly below \lstar.
Many attempts have been made to pin down \lfha\ and its evolution with
redshift from $z\approx 0$ 
\citep[e.g.][]{Gallego1995}, through intermediate-$z$
\citep{Yan1999,Hopkins2000,Tresse2002,Sobral2009,Shim2009}, 
with the first $z=2$ limits placed by 
\cite{Bunker1995}.
It is only in recent years that $z=2$ surveys have been fruitful, with
the most significant study being that of
\citet[][hereafter G08]{Geach2008}. 
However, this study was a wide, shallow survey designed to 
find the bright objects and determine \lstar, but does not permit an estimate
of the faint-end slope. 

With the wide-field (7\farcm5$\times$ 7\farcm5) {\em HAWK-I} instrument 
\citep{Pirard2004,Casali2006}
at {\em ESO-VLT} we have obtained the deepest
narrowband \halpha\ observations to date as part of a programme to study 
\halpha\ and \lya\ emitting galaxies from a single volume at 
$z\sim 2.2$ (Hayes et al., submitted).
These \halpha\ data alone enable us to study the faint-end of 
\lfha\ at unprecedented depths, tightening constraints on the overall
\lfha, and providing the content of this \emph{Letter}.
In Sect.~\ref{sect:data} we briefly describe the data, reduction, and 
selection;
in Sect.~\ref{sect:results} we present the $z\sim 2$ \halpha\ luminosity 
functions we derive and the constraints on \sfrd; 
and in Sect.~\ref{sect:summary} we briefly summarise.
Throughout we adopt a cosmology of 
$(H_0, \Omega_{\mathrm M}, \Omega_{\Lambda }) = 
(70~\mathrm{km~s}^{-1}~\mathrm{Mpc}^{-1}, 0.3, 0.7)$, 
the SFR(\halpha) calibration of 
\citet{Kennicutt1998}, 
derived for a Salpeter initial mass function in the 
mass range $(M_\mathrm{lo}, M_\mathrm{hi}) = (0.1,100)M_\odot$ and solar 
abundances,
and the $AB$ magnitude system 
\citep{OkeGunn1983}.

\section{The data\label{sect:data}}

\subsection{Observations and reduction}

A field in the GOODS South 
\citep{Giavalisco2004}
was selected to maximise the depth and quality of the auxiliary data.
The central pointing was $\alpha=$03:32:32.88; $\delta=$--27:47:16; 
p.a.$=-44\deg$.
The field was observed in service mode between 08 September and 21 August 2008 
on a single dithered pointing. 
The {\em NB2090} filter ($\lambda_\mathrm{c}= 2.095$\micron; 
$\Delta\lambda =0.019$\micron) was used to isolate
emission line candidates, using a total integration time of 60\,000~seconds. 
For the continuum we obtained 7\,500~s in {\em HAWK-I} \ks\ band,
which we combined with the publicly available {\em ISAAC} \ks\ data from the
GOODS/ESO Imaging Survey. 

Full details of the reduction process are beyond the scope of 
this {\em Letter} and will be presented in a forthcoming article 
(Hayes et al. 2010). For brevity, the \emph{EsoRex HAWK-I} 
pipeline was used on the individual frames for bias-subtraction, 
flat-fielding, and subtraction of the sky with temporally adjacent image 
pairs.
Custom scripts were then used to mask cross-talk artifacts, register and 
co-add the individual frames. 
We modelled the point spread function (PSF) in the final image and 
determined a seeing of 0.89\arcsec.
To estimate the limiting magnitude we added artificial point-sources (full width
at half maximum set to the measured seeing) to the images 
with the {\tt addstar} task in {\tt noao/iraf} and tested their recovery using 
{\tt SExtractor} 
(\citealt{BertinArnouts1996}; 
see Sect.  \ref{sect:selection}).  
It should be noted that the limiting magnitude for extended objects will be 
somewhat shallower, but also that at $z=2.2$, one seeing disc corresponds to a 
physical scale of 7.4~kpc.
Using {\tt SExtractor mag\_auto} magnitudes we determined a $5\sigma$ 
limiting magnitude of 24.6 in \emph{NB2090}. 
By computing the product of this flux density and $\Delta\lambda_{NB2090}$,
this corresponds to a line flux of $6.85\times 10^{-18}$~erg~s$^{-1}$, if all 
the \emph{NB2090} flux comes from \halpha\ and falls at the peak of the filter
throughput. This assumption, valid for perfect top-hat bandpasses, holds 
well for our bandpass, which has steep edges by narrowband standards: the 
full width at 80\% transmission is over 80\% the FWHM. 
This limiting flux corresponds to SFR=$1.9M_\odot$~yr$^{-1}$. 
The \ks\ $5\sigma$ limit is 25.4.

\subsection{Selection of H-alpha emitting candidate
	galaxies\label{sect:selection}}

\begin{figure}
\centering
\includegraphics[width=9cm]{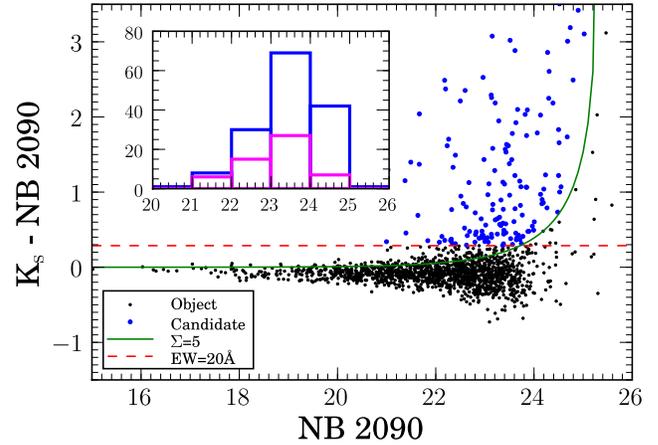}
\caption{The selection function for candidate \halpha\ emitters. Black dots
represent all the objects detected in the \emph{NB2090} image. The red dashed line
shows the colour cut of \ewha$= 20$~\AA, and the green solid line
the limit of $\Sigma=5$. The blue points show the selected candidate
galaxies. The inset histogram shows the narrowband magnitude distribution of 
all candidate emission line galaxies in blue, and confirmed 
$z\approx2.2$ \halpha\ emitters after the rejection of interlopers 
(see text) in magenta.}
\label{fig:selection}
\end{figure}

Source detection was performed in the on-line image using {\tt SExtractor}, 
where we required a minimum of 12 contiguous pixels 
(plate-scale=0\farcs 106~px$^{-1}$) 
to reside above the background noise by a signal--to--noise ($S/N=3$).
Aperture-matched \ks\ photometry was performed using ``double image" mode.
We selected emission line candidates based upon two criteria, the first of 
which was a minimum rest-frame equivalent width, \ewha$=20$\AA.
Furthermore, in order to prevent over-contamination from $K_\mathrm{s}$-faint 
objects scattering
into the colour-selection region, we required the narrowband flux to be a factor 
of $\Sigma=5$ greater than the noise in the continuum image. 
Figure~\ref{fig:selection} shows the colour-magnitude selection diagram, 
including the cuts in \ewha\ and $\Sigma$.

\begin{figure*}[t]
\centering
\includegraphics[width=8cm]{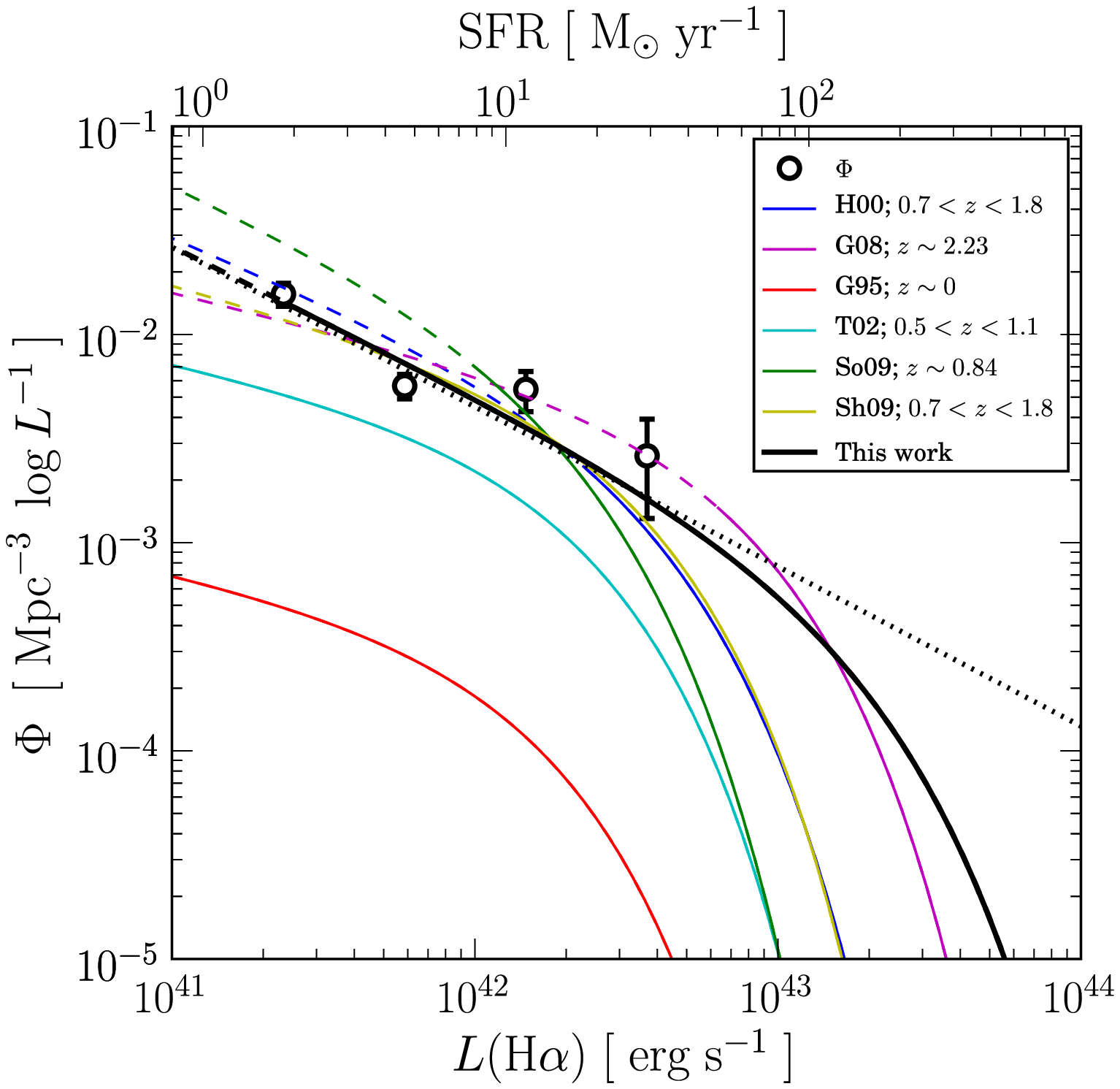}
\includegraphics[width=8cm]{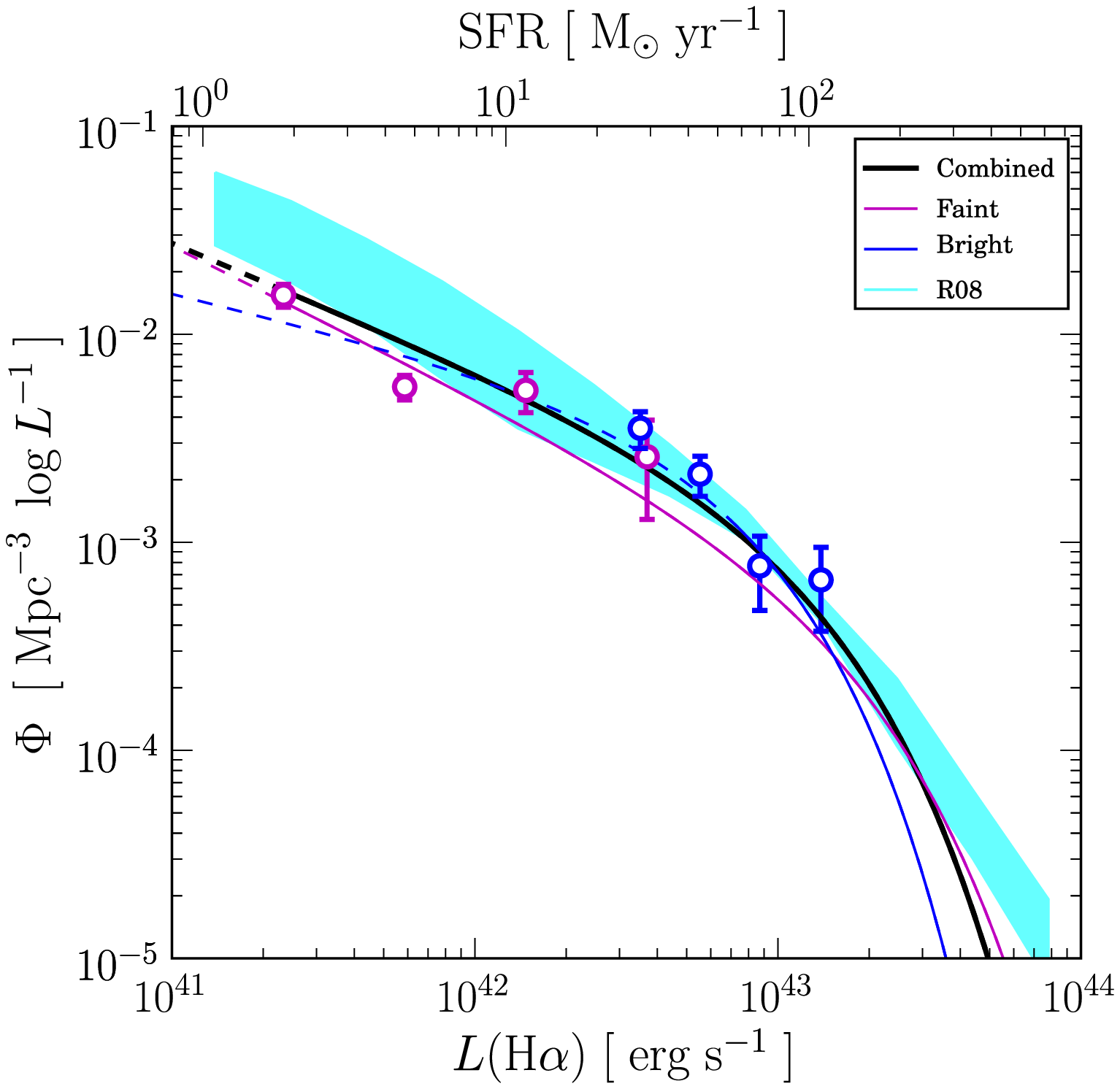}
\caption{\emph{Left}: \halpha\ luminosity functions. The black points show 
the bins created from \emph{HAWK-I} data obtained for our programme with 
error-bars derived from Poisson statistics and Monte Carlo incompleteness 
simulations. 
The best-fit Schechter function is shown by the black solid line, while the  
black dotted line represents the best fitting simple power-law.
Other lines show various luminosity functions from the 
literature, scaled to our cosmology.
H00=\cite[][all points]{Hopkins2000},
G08=\cite{Geach2008},
G95=\cite{Gallego1995},
T02=\cite{Tresse2002},
So09=\cite{Sobral2009}.
Sh09=\cite{Shim2009}.
Dashed vs. solid representation shows the $3\sigma$ limiting survey depth.
\emph{Right}: The complete luminosity function combining these \emph{HAWK-I}
data with the points from G08. The shaded cyan area shows the 
\lfha\ derived by \citet{Reddy2008} from the modelling of $<z> = 2.3$ 
colour-selected galaxies.  }
\label{fig:lf}
\end{figure*}

After inspecting each candidate by eye, we found 152 objects 
that could potentially be emitters of any emission line between Pa$\alpha$ at 
$z\approx 0.11$ and \lya\ at $z\approx 16$.
We then cross-correlated our candidates with the merged $z-$ and \ks-selected 
\emph{GOODS-MUSIC} catalogue of 
\citet{Santini2009}, 
finding 143 matches. 
The objects for which counterparts were not detected all appear to be 
genuine narrowband-excess objects, but with equivalent width lower limits so 
high that their stellar continua are too faint to be detected in the 
\emph{ISAAC} \ks\ or \emph{ACS} $z-$band images. 
\emph{GOODS-MUSIC} provided us with spectral energy distributions (SEDs)
between the $U-$band and \emph{Sptizer}-MIPS 24\micron.
To assemble the final catalogue we first examined the spectroscopic
redshift measurements in the \emph{GOODS-MUSIC} catalogue. If the 
spec-$z$ had a quality flag of ``very good" or ``good" and was consistent 
with the redshift interval defined by the \emph{NB2090} bandpass 
($2.178 < z < 2.207$) the galaxy was included (1 galaxy).
Using the same quality criterion, we excluded objects if they fell outside 
that $z$ range (26 galaxies).
This ratio of 26 galaxies excluded versus one included may superficially appear to 
show the selection method in a rather negative light, but the bias can easily be 
explained by considering the selection methods of the many studies that followed
up the GOODS imaging spectroscopically. 
We searched the spectroscopic catalogues for all galaxies in \emph{GOODS-MUSIC} 
with spec-$z$ in the range covered by $FWHM_{NB2090}$ and spec-$z$ flags of 
good or better, finding only a single object that could be detected by our 
survey: the one we do find, followed up from $BzK$ selection.
In contrast the catalogue is rife with spectra targeting the $z=0-2$ domain 
and Lyman-break galaxies at $z>3.5$, and the combined catalogue is heavily 
biased against $z=2-3$.
We find typically as interloping lines: $z\sim 3.2$ [\ion{O}{iii}]; low-$z$
Pa$\alpha$, Pa$\beta$ and lines in the higher order Brackett series; and 
$z\sim 0.28$ [\ion{Fe}{ii}] emitters, although a handful of more unknown or 
interesting lines are also recovered.
The number of emission line candidates is of sufficient interest to warrant 
further examination, which will be the topic of a forthcoming paper.
For objects with no spec-$z$ or an uncertain flag, we used the \emph{Hyper-z} 
photometric-redshift code 
\citep{Bolzonella2000},
modified to include nebular continuum and lines 
\citep{SchaererdeBarros2009}.
Here we selected objects that had $1\sigma$ errors on their phot-$z$ 
consistent with $z=2.19$.
Combined with the spectroscopically confirmed objects this gave us 55 
$z\approx 2.2$ \halpha-emitters.
For security, we tested our galaxies against the $BzK$ colour criterion of 
$(B-z)-(z-K) \ge -0.2$ 
\citep{Daddi2004}
to insure that our objects would be classified as star-forming objects at 
$z\ge 1.4$. 
Only four of our 55 objects do not satisfy this criterion, but when examining
the $BzK$ colours of the entire narrowband-excess sample (143) we found an 
additional 7 objects that do. 
We note also that the $BzK$ criteria neglect the fact that the $B$ and 
$K$ photometry may be contaminated by the \lya\ and \halpha\ emission lines, 
respectively. 
Since the number of objects that are classified/declassified by $BzK$ colours 
are (\emph{i.}) so similar, (\emph{ii.}) only a small ($\sim 10$\%) fraction of 
our sample, and (\emph{iii.}) possibly contaminated by lines, we opt to 
leave our phot-$z$ selected sample unchanged. 
Finally, we cross-correlated our sample against the 1~Mega-second 
\emph{Chandra} Catalogue 
\citep{Giacconi2002,Rosati2002},
but no objects in which the \halpha\ production is obviously dominated 
by an active nucleus were found, including the objects missed by broadband 
cross-correlation.
Our final catalogue comprises 55 objects.

\section{Results and Discussion\label{sect:results}}

\subsection{The redshift 2.2 H-alpha luminosity function\label{sect:lfha}}

All \halpha\ photometry is corrected for the contribution of 
[\ion{N}{ii}]$\lambda\lambda6548,6583$\AA\ lines.
In local starbursts the [\ion{N}{ii}]/\halpha\ ratio varies strongly with 
metallicity ($Z$) between $\sim 0$ and 0.6, and 
[\ion{N}{ii}]/\halpha\ may plausibly be estimated from \lha\ through the 
luminosity--metallicity relationship.
However a significant offset in the $L-Z$ relationship is found at 
$z\sim 2$ 
\citep{Erb2006a},
and an application of the local relationship at $z\approx 2$ would be insecure. 
Furthermore, the 
\cite{Erb2006a}
study 
is based on galaxies an order of magnitude brighter than those found here, and 
we dare not extrapolate to these luminosities. 
Thus we apply a conservative single correction of 
[\ion{N}{ii}]/\halpha$= 0.117$, derived by averaging over the faintest four 
bins in the data of 
\cite{Erb2006a}.

\begin{figure}[t!]
\centering
\includegraphics[width=9cm]{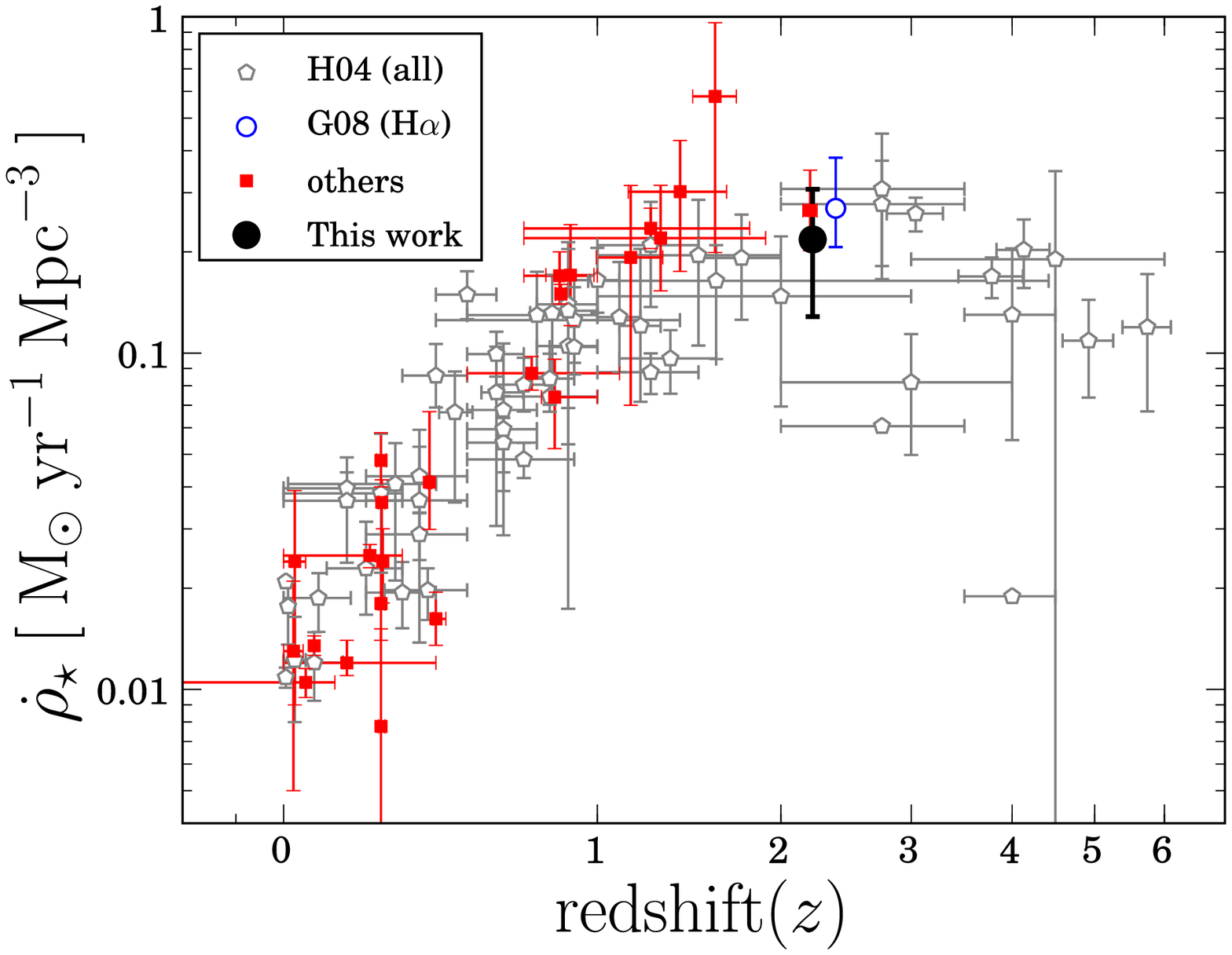}
\caption{The cosmic star-formation rate density. Open grey pentagons show all of the 
dust-corrected points compiled by 
\cite{Hopkins2004} 
that were derived from anything other than \halpha. 
Filled red squares show dust-corrected points from 
\cite{Shim2009},
all of which are \halpha-based.
The black point is the measurement from this study and the open blue point 
is the result 
from G08, artificially shifted to higher-$z$ to allow it to be distinguished 
from our own.  }
\label{fig:sfrd}
\end{figure}

\lfha\ is created by binning all selected objects by luminosity, with 
errors derived from the standard Poisson statistic and Monte Carlo simulations
of incompleteness. 
We fit the Schechter function using a standard $\chi^2$ minimiser, determining
errors from 1\,000 re-drawn Monte Carlo realisations. 
The best-fit Schechter parameters are presented in Table~\ref{tab:lfpars} 
and are shown graphically with other determinations from the literature in 
Fig.~\ref{fig:lf}.

\begin{table}
\begin{minipage}{\columnwidth}
\caption{Luminosity function parameters}
\label{tab:lfpars}      
\centering                   
\renewcommand{\footnoterule}{}  
\begin{tabular}{c c c c c}       
\hline\hline                 
Param. & Unit
	& HAWK-I all\footnote{3 Schechter parameters: \emph{HAWK-I} data only.} 
	& HAWK-I $\alpha$ \footnote{$\alpha$ only: \emph{HAWK-I} data only.} 
	& Combined \footnote{3 Schechter parameters: \emph{HAWK-I} data + G08 points.} \\    
\hline                        
 $\log L_\star$    & erg~s$^{-1} $ 
 				& $43.22\pm 0.56$ & ...             & $43.07\pm  0.22$ \\      
 $\log \phi_\star$ & Mpc$^{-3}$                
 				& $-3.96\pm 0.94$ & ...             & $-3.45\pm  0.52$ \\
 $\alpha$          & ...                       
 				& $-1.72\pm 0.20$ & $-1.77\pm 0.21$ & $-1.60\pm  0.15$ \\
\hline                                   
\end{tabular}
\end{minipage}
\end{table}

The brightest galaxy in our sample has a luminosity of 
$3.5 \times 10^{42}$~\lumcgs, which is a factor of 4.5 fainter than the value of
\lstar\ we derive, and even allowing for the error on \lstar, we clearly 
are not probing this luminosity domain. 
This galaxy is a factor of 2 fainter than the 
$L_\star = 6.7\times 10^{42}$~\lumcgs\ derived by 
G08 and demonstrates the inadequacy of such a small survey to 
constrain all Schechter parameters. Indeed, the statistical error on the 
degenerate parameter \phistar\ spans an entire dex and it is apparent 
that the only parameter we can reliably constrain is $\alpha$, since all our 
\halpha\ emitters have luminosities that place them firmly in the power-law 
region of the $LF$. 
For security, we proceed to examine the faint-end only by fitting of a simple 
power-law, using exactly the same Monte Carlo realisation data as previously.
Our derived faint-end slope is found to be $-1.77\pm 0.21$, also
presented in Table~\ref{tab:lfpars}, and exhibits a near-identical 
mean and errorbar. 

Some authors find a steepening in the faint end slope of \lfha\ moving from 
$z=0$ to $z\gtrsim 1$ with similar results found in UV-selected catalogues at 
$z=2$ to 3 
\citep{ReddySteidel2009}
and $z=6$
\citep{Bouwens2007}.
In addition, 
\cite{Hopkins2000}
find a steepening in $\alpha$ for \halpha\ emitters at $z \sim 0.7$ to 1.8 
by slitless spectroscopy with the \emph{HST/NICMOS},
with $\alpha=-1.60$ or 1.86 depending on the selection criteria. 
However, 
\cite{TrentiStiavelli2008}
note that for such small field--of--view studies, an artificial steepening
of the faint-end slope may manifest itself if under-dense regions are targeted. 
Examining the redshift distribution of galaxies in the whole GOODS field
using both spec-$z$ and phot-$z$ reveals $z=2.2$ to correspond to neither a 
significant over- nor under-density. 
After deriving $\alpha = -1.77 \pm 0.21$, we are able to confirm this increase
in the relative abundance of lower-luminosity galaxies at epochs of peak
cosmic star-formation. 
This confirmation comes at the $3\sigma$ level with respect to 
\cite{Gallego1995} 
($z=0$), who find $\alpha=-1.35$, and our results are consistent with no 
evolution within $1\sigma$ to $z=0.84$ 
($\alpha=-1.65$; \citealt{Sobral2009})
and $0.8 \lesssim z \lesssim 1.7$
($\alpha=-1.86$; \citealt{Hopkins2000}).

\subsection{Combined \lfha\  and the star-formation rate density\label{sect:sfrd}}

As stated above, our data are not sufficient to constrain the brighter end of 
\lfha.  To address this we take the \lfha\ data-points from G08
and combine them with our own. 
We extract the $LF$ data-points, which have been corrected for the 
[\ion{N}{ii}] using the $z\sim0$ estimate of 
\cite{KennicuttKent1983}, 
and reapply the $z\sim2$ correction of 
\cite{Erb2006a} 
for consistency with our own points.
The resulting $LF$ can be seen in the \emph{right} panel of Fig.~\ref{fig:lf} 
with the best-fitting Schechter parameters, derived following exactly the same
method as in Sect.~\ref{sect:lfha} in Table~\ref{tab:lfpars}.
We also show an estimate of the 
observed (i.e. dust-uncorrected) \lfha\ at $<z> = 2.3$ from 
\citet{Reddy2008},
inferred from colour-selected samples by equating dust-corrected 
SFR(UV) with SFR(\halpha) and re-applying the effects of dust attenuation to
\halpha.
The observed and inferred \lfha\ agree well in general, particularly 
around \lstar, with only a mild deviation at the faint and (unsampled) bright ends.

In order to estimate the cosmic star-formation rate density, we need to
correct our \halpha\ fluxes for dust attenuation.
To do this we re-fit the SEDs of all the \halpha-emitters, fixing their 
redshift to 2.19. 
Ideally we would correct our luminosity function bin by bin, but no 
obvious correlation is observed between \lha\ and \av. Thus we adopt the 
average \av\ for the sample of $1.19 \pm 0.55$, assuming the attenuation law of
\cite{Calzetti2000}.
We do not employ a correction factor for the effects of differential 
attenuation of stellar and nebular light, since at $z\sim 2$ 
\cite{Erb2006b} find SFR$(\mathrm{UV})$ equivalent to SFR$(\mathrm{H}\alpha)$ 
without this factor.
\av$=1.19$ equates to $A_{6563\mathrm{\AA}}=0.977$ for the
fainter bins in \lfha, identical to the assumption of 1~mag used by 
G08
at the brighter end, and we correct for 1 mag of attenuation throughout.
It is noteworthy that 
\citet{Reddy2008} 
find a maximum likelihood \ebv\ of 0.12 ($A_{6563\mathrm{\AA}}=0.4$, Calzetti 
dust) for \emph{BX}--selected galaxies at $<z>=2.3$.
The significantly different extinction determined from \halpha\ selection may
be due to the very different selection functions: indeed \ewha\ is independent 
of dust extinction, whereas \emph{BX} selection cuts the multi-broadband sample
based on $U-G$ colours, and thus it is not unreasonable to infer that galaxies 
with very different extinctions survive the respective cuts. 
The errors derived on the Schechter parameters are strongly correlated, and 
thus we compute \sfrd\ from each of the raw Monte Carlo realisations 
previously described. For consistency with 
\cite{Hopkins2004} 
we integrate over the luminosity range 
$37\le \log (L_{\mathrm{H}\alpha} /\mathrm{erg}~\mathrm{s}^{-1})\le 47$,
where we find a value of 
$\rho_\mathrm{L} = (2.72 \pm 1.14)\times 10^{40}$~erg~s$^{-1}$~Mpc$^{-3}$
and 
$\dot{\rho}_{\star} = (0.215\pm 0.090)$\sfrvol.
This however represents a significant extrapolation over luminosity, and 
we also present the same integration performed over the range covered by our 
bins in total \lfha: 
$41.5\le \log (L_{\mathrm{H}\alpha} /\mathrm{erg}~\mathrm{s}^{-1})\le 43.1$.
This value, much less dependent on the Schechter parameter fitting,  results in 
$\rho_\mathrm{L} = (1.76 \pm 1.02)\times 10^{40}$~erg~s$^{-1}$~Mpc$^{-3}$
and
$\dot{\rho}_{\star} = (0.139 \pm 0.081)$~\sfrvol. 
We show the result in Fig.~\ref{fig:sfrd}, along with two 
compilations of $\dot{\rho}_{\star}$
from the literature. 
We include the dust-corrected \halpha\ survey data compiled by 
\cite{Shim2009},
and points from the  
\cite{Hopkins2004} 
compilation based upon many different observations where we have excluded all 
determinations based upon \halpha.
We also indicate separately the point of G08, which has been artificially 
moved in redshift to separate it from our own. 
Our point agrees well with that of G08, which is initially surprising,
given that we obtain a significantly steeper faint-end slope and marginally higher
\lstar.
The culprit behind the similar \sfrd\ yet differing (\lstar,$\alpha$) is 
naturally the overall normalisation of the luminosity function, \phistar, which 
we calculate to be a factor of 4 lower. 
Furthermore the errorbar produced by our analysis is actually \emph{larger}
than that published by G08,
despite our $LF$ extending an order of magnitude fainter in luminosity. 
These errorbars, however, are statistical errors that result from averaging 
over many Monte Carlo realisations, and our fitting engine includes an 
additional parameter ($\alpha$) -- fixing one parameter introduces a 
systematic error that is not accounted for by Monte Carlo. 
Had we locked $\alpha$ to its best-fit value and fitted \lstar\ and 
\phistar as usual, our statistical errorbar on \sfrd\ would be a factor of 3 smaller.
Finally it is possible that our selection criterion of \ewha$>20$\AA\ could 
cause us to underestimate the total luminosity density. To assess the impact of
this we examine the \ewha\ distribution of our sample and fit an exponential of 
the form $N \propto \exp (EW/EW_0)$, finding 
$EW_{\mathrm{H}\alpha,0} = 115$\AA.
Assuming the two extremes of \ewha\ to be independent of \lha\ and also 
of the continuum luminosity, we find that our selection criterion misses 
between 1.3 and 16\% of the integrated luminosity.
We also examine the \ewha\ distribution of nearby galaxies in the 
\emph{Sloan Digital Sky Survey}, selecting the complete spectroscopic sample 
at $z<0.1$, and find that removal of galaxies with \ewha$<20$\AA\ cuts 
19\% of the \halpha\ luminosity density.

\section{Summary\label{sect:summary}}

We have used the new \emph{HAWK-I} instrument mounted at \emph{ESO-VLT} to 
carry out a blind, narrowband survey for \halpha\ emitting galaxies at a 
redshift of 2.2. 
This is the deepest unbiased survey carried out at this redshift to date and 
enables us to estimate the faint-end of the \halpha\ luminosity function, a 
parameter hitherto merely assumed at this cosmic epoch. 
The target is the GOODS-South field, offering us a rich, deep multi-wavelength
ancillary data-set, in which we find 55 \halpha\ emitters.
\begin{enumerate}
\item {We construct a luminosity function from the \emph{HAWK-I} data which we 
find is well fit by a Schechter function with parameters of 
$\log L_\star    = (43.22 \pm 0.56)$~\lumcgs,
$\log \phi_\star = (-3.96 \pm 0.94)$~\permpc,
$\alpha          = (-1.72 \pm 0.20)$.
Fitting a single power-law component to all the LF points gives a very similar 
value of $\alpha = (-1.77 \pm 0.21)$.
This confirms the steepening of the faint-end of the galaxy 
luminosity function out to $z=2$, which is predicted and observed at 
other wavelengths.}

\item {We combine our luminosity function bins with those from the much wider 
but shallower survey of G08, which
yields the best-sampled \halpha\ luminosity function at this redshift. 
This is also well described by a Schechter function with the parameters of 
$\log L_\star    = (43.07\pm 0.22)$~\lumcgs,
$\log \phi_\star = (-3.45\pm 0.52)$~\permpc,
$\alpha          = (-1.60\pm 0.15)$.}

\item {We apply corrections for dust attenuation derived from modelling of the 
stellar spectral energy distributions and integrate the combined luminosity 
function to obtain an instantaneous cosmic star-formation rate density of 
$(0.215\pm 0.090)$~\sfrvol. This provides the most robust emission-line estimate 
to date at this redshift.}

\end{enumerate}

\begin{acknowledgements}
M.H. \& D.S. are supported by the Swiss National Science Foundation. 
G\".O. is Royal Swedish academy of Sciences Research Fellow, supported by a grant
from the Knut and Alice Wallenberg Foundation and acknowledges support from the
Swedish Research Council. 
We thank Naveen Reddy for making specific \lfha\ realisations available,
Hyunjin Shim for sharing the compilation of literature $\dot{\rho}_{\star}$
values. We thank Claudia Scarlata for useful discussions.
Finally we  thank the anonymous referee for thoroughly dissecting the manuscript
and suggesting numerous improvements. 
\end{acknowledgements}

\bibliography{half.bib}

\end{document}